\documentclass[prd, twocolumn, nofootinbib]{revtex4}

\usepackage{epsfig} 

\newcommand{\beq}{\begin{equation}}
\newcommand{\eeq}{\end{equation}}
\newcommand{\beqa}{\begin{eqnarray}}
\newcommand{\eeqa}{\end{eqnarray}}
\newcommand{\om}{\Omega_m}

\newcommand{\ls}{\mathrel{\raise0.27ex\hbox{$<$}\kern-0.70em \lower0.71ex\hbox{{
$\scriptstyle \sim$}}}}

\begin{document} 

\title{Redshift Distortions as a Probe of Gravity} 
\author{Eric V. Linder}
\affiliation{Berkeley Lab, University of California, Berkeley, CA 94720, USA} 

\date{\today}

\begin{abstract} 
Redshift distortion measurements from galaxy surveys include sensitivity 
to the gravitational growth index distinguishing other theories from 
Einstein gravity. 
This gravitational sensitivity is substantially free from uncertainty 
in the effective equation of state of the cosmic expansion history. 
We also illustrate the bias in the traditional application to matter 
density determination using $f=\om(a)^{0.6}$, and how to avoid it.
\end{abstract} 


\maketitle

\section{Introduction \label{sec:intro}}

Measurements of growth of cosmic structure open windows on probing the theory 
of gravity and, in combination with observations of the expansion history, 
can distinguish between a physical dark energy and extended gravity as 
an explanation for the accelerating expansion of the universe.  As pointed 
out by \cite{guzzo}, the pattern of redshift distortions around large
scale structure over cosmic time could offer such a probe of growth history.  
Recent developments such as the new VVDS redshift survey \cite{guzzo} 
and analysis of existing data sets \cite{ross} motivate examination 
of this prospective probe.  This technique has the advantage of 
using galaxies purely as markers, ameliorating 
selection effects, and requiring only moderate resolution spectroscopy 
\cite{guzzo}. 

Redshift distortions are caused by velocity flows induced by gradients 
of the gravitational potential, with the gravitational potential 
evolving due to both the growth under gravitational attraction (given 
by general relativity or some other theory of gravity to be considered) 
and the dilution of the potentials due to the cosmic expansion 
(characterized by the effective equation of state).  
Because the velocity field involves one less spatial derivative of the 
potential than the density field does, the signal persists to larger, 
more linear scales, and offers the hope of serving as a cosmological 
and gravitational probe. 

We investigate the sensitivity of this probe in \S\ref{sec:sens}, 
calculating its leverage on distinguishing the theory of gravity and the 
equation of state.  Its complementarity with expansion probes is considered 
in \S\ref{sec:compl}.  In \S\ref{sec:om} we demonstrate how the traditional 
use of redshift distortions to determine the matter density becomes biased 
if the incorrect form is assumed.

\section{Sensitivity \label{sec:sens}}

The redshift distortion parameter observed through the anisotropic pattern 
of galaxy redshifts on cluster scales is \cite{kaiser} 
\beq 
\beta=\frac{1}{b}\frac{d\ln D}{d\ln a}=f/b, 
\eeq 
where $b$ is the bias between galaxies and the total matter, and $D$ 
is the linear theory growth factor at expansion scale factor $a$.  
Considering the gravitational growth index formalism of \cite{groexp}, 
\beq 
f=\om(a)^\gamma, 
\eeq 
where $\om(a)$ is the matter density as a fraction of the total 
energy density at scale factor $a$ and $\gamma$ is the gravitational 
growth index.  Note that, as it was designed for, the growth index 
formalism separates out the two physical effects on the growth of 
structure and the redshift distortion: $\om(a)$ involves the expansion 
history and $\gamma$ focuses on the gravity theory.  We will see that 
this is particularly beneficial in the present case.

\begin{figure}[!htb]
\begin{center} 
\psfig{file=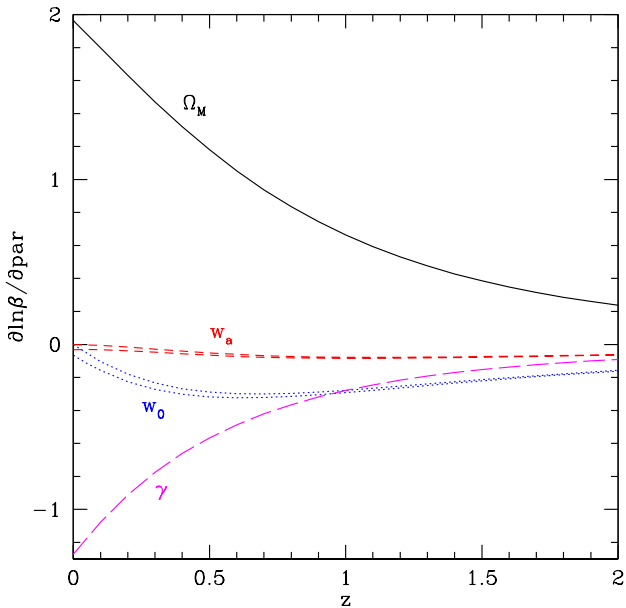,width=3.4in} 
\caption{The sensitivities of the redshift distortion $\beta$ to 
variations in the cosmological parameters.  The unmarginalized 
uncertainty in a parameter is given by $\sigma(p)=\sigma_{\ln\beta}/|\partial 
\ln\beta/\partial p|$.  For example, for a 10\% precision on $\beta$ 
at $z=0.8$, the uncertainty on $w_0$ would be $0.1/0.3\approx 0.3$. 
The upper curves for $w_0$ and $w_a$ show the sensitivity when the 
slight dependence of $\gamma$ on $w(z)$ is neglected. 
}
\label{fig:betasens} 
\end{center} 
\end{figure}

To investigate the leverage of redshift distortion measurements at 
various redshifts on the cosmological parameters, we use a Fisher 
information matrix calculation to propagate uncertainties in $\beta$ 
to the estimations of the present matter density $\om$, the gravitational 
growth index $\gamma$, and the effective 
equation of state parametrized by $w(a)=w_0+w_a(1-a)$, with $w_0$ the 
present effective dark energy equation of state, or pressure to energy 
density, ratio and $w_a$ a measure of its time variation.  Initially 
we consider the bias parameter $b(a)$ to be well known.  
Figure~\ref{fig:betasens} shows as a function of redshift the Fisher 
partial derivatives giving the sensitivities. 

The gravitational growth index has substantial impact on the redshift 
distortion, with $\beta$ more sensitive to the growth index $\gamma$ 
than to the equation of state variables $w_0$, $w_a$, 
especially at low redshifts.  This is furthermore where 
measurements can be made most precisely, so this suggests that redshift 
distortions offer a promising tool for investigating gravity. 
Sensitivity to the equation of state parameters is modest, so redshift 
distortions do not serve as a strong dark energy expansion probe.  However, 
this insensitivity combined with the very different shapes of the $\gamma$ 
and $w_0$, $w_a$ curves mean that there is not much degeneracy between the 
gravity and the expansion history parameters -- a promising sign. 

Suppose we consider an optimistic scenario of 10\% measurements of 
$\beta$ in 20 redshift bins from $z=0.1-2$.  Note that current state of 
the art is a 37\% measurement of $\beta(z=0.77)$ \cite{guzzo}, 12\% 
measurement (using additional clustering information) of $\beta(z=0.55)$ 
\cite{ross}, and 18\% measurement of $\beta(z=0.15)$ \cite{hawkins}. 
Additionally we include 
CMB constraints of 0.7\% in the reduced distance to last scattering, 
{\`a} la Planck, which are independent of $\gamma$.  The unmarginalized 
uncertainty in the determination of the gravitational growth index is 
$\sigma(\gamma)=0.048$.  (Note that the difference between general relativity 
and the extradimensional DGP braneworld gravity scenario is 
$\Delta\gamma=0.13$.)  If we marginalize over $\om$, keeping to the 
$\Lambda$CDM $w=-1$ scenario, then $\sigma(\gamma)=0.061$, and if we also fit 
for a constant $w$ then this becomes 0.067, and finally if we marginalize 
over a time varying $w(z)$ as well, i.e.\ $w_0$ and $w_a$, then the 
uncertainty is 0.077.  

The increase of 61\% from unmarginalized to fully 
marginalized estimation is remarkably small. (By contrast, uncertainty in 
$w_0$ would increase by a factor 12.)  This reflects that $\gamma$ 
predominantly tests the gravity theory independent of the expansion 
history. (Indeed, $\delta\gamma\approx\delta w/20$.)  
The growth index estimation is also quite distinct from 
uncertainties in the matter density, with the correlation coefficient 
between $\gamma$ and $\om$ only 0.12. 

While from Fig.~\ref{fig:betasens} we see that the main leverage on 
$\gamma$ arises from low redshift, the higher redshift measurements 
help not only through more statistics but in breaking the residual 
degeneracies.  Thus, to achieve these constraints on $\gamma$ one 
must succeed in precision measurements of redshift distortions to $z>1$.  
If the data only goes out to $z=1$ then the marginalized uncertainty on 
$\gamma$ increases to 0.16, while preserving the original statistical power 
(e.g.\ by improving the precision by $\sqrt{2}$ of the now half as many 
measurements) yields an uncertainty of 0.12, compared to 
the original ($z=0.1-2$ data) value of 0.077.

\section{Complementarity \label{sec:compl}} 

However, if we employ the redshift distortion method in complementarity 
with an expansion history probe such as Type Ia supernovae (SN), then we 
can obtain tight constraints on the gravitational growth index $\gamma$ 
with a more modest dataset in $\beta$, extending only over $z<1$.  
(Note that using baryon acoustic oscillations as the distance probe 
could be somewhat problematic, as they employ the same density field 
as the redshift distortions.)  Combining redshift distortion 
measurements from $z=0.1-1$, plus CMB data, with SN observations from 
$z=0.1-1.7$ as 
from the future SNAP satellite (recall SN are purely an expansion probe, 
with no leverage on $\gamma$), we can fit for the growth index to 0.057 
with negligible change in the SN determination of $w_0$, $w_a$.  Thus the 
two methods work in parallel to test both gravity and dark energy. 

\begin{figure}[!htb]
\begin{center}
\psfig{file=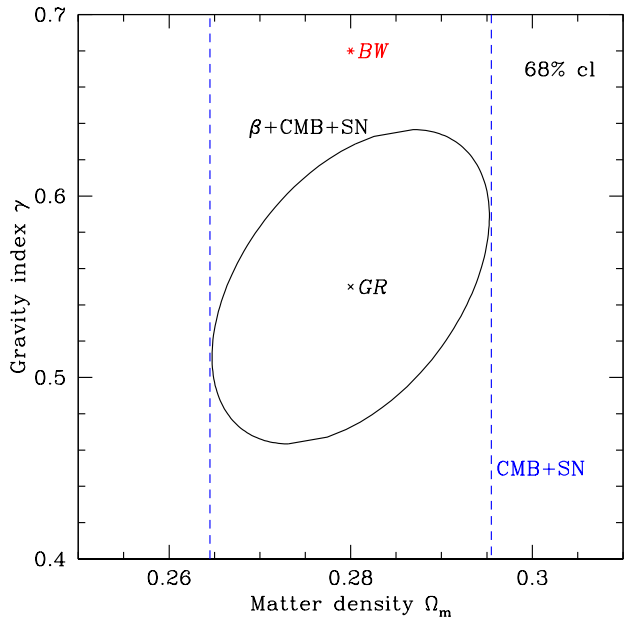,width=3.4in}
\caption{Complementarity of measurements of the redshift distortion 
factor $\beta$, here ten data points from $z=0.1-1$ at 10\% precision, 
with CMB and supernovae measurements of the expansion history, provides 
an appreciable test of gravity through the growth index $\gamma$.  While 
the expansion probes are blind to $\gamma$ (vertical band), use of 
redshift distortion 
data can distinguish the Einstein value $\gamma=0.55$ (shown by the 
black x) from braneworld gravity having $\gamma=0.68$ (red star). 
}
\label{fig:omgam}
\end{center}
\end{figure}

Figure~\ref{fig:omgam} shows the constraints in the $\om$-$\gamma$ plane, 
marginalizing over the equation of state parameters, with the combination 
of redshift distortion $\beta$, CMB, and SN 
able to distinguish deviations in gravitational index.

Adding new parameters to describe the galaxy bias makes the situation 
more challenging.  Because of the constraints from the distance 
data, the bias acts 
primarily to increase the uncertainty on $\gamma$.  Considering two new 
parameters to describe the bias, $b(a)=b_0+b_1(1-a)$, one requires priors 
on these parameters of 0.03 (0.05) to prevent the estimation of $\gamma$ 
from increasing by more than 20\% (40\%).  
For future surveys one can hope to limit the galaxy bias uncertainty 
through use of the higher order correlation function data present in 
the survey, and simulations 
taking advantage of that the velocity field probes larger, more linear 
scales than the density field.  

Moreover, note that an additional low 
redshift ($z\approx0.05$) measurement helps 
considerably because there the growth factor involves nearly the present 
matter density (well determined by complementarity with the distance 
information), and the growth index, thus reducing the uncertainty on $\gamma$. 
The very optimistic situation of 5\% measurements of redshift distortions 
from $z=0.1-1$, including at $z=0.05$ and priors of 0.03 (0.05) on 
the bias parameters, with distance data, could allow a seven parameter fit, 
delivering knowledge of $\gamma$ to 0.041 (0.047).

\section{Matter Density (Mis)Estimation \label{sec:om}}

Traditionally, redshift distortion measurements have been used to 
determine the matter density (see \cite{ballinger,matsu,hawkins} and 
references 
therein).  As seen in Fig.~\ref{fig:betasens}, $\beta$ has considerable 
sensitivity to $\om$.  However, the correct form for growth factor must 
be employed; the approximation $f=\om(a)^{0.6}$ from 
1980 \cite{pje80} is appropriate only for a universe with purely matter, 
of which a fraction $\om$ clusters.  A highly accurate 
growth index form is $f=\om(a)^\gamma$, with $\gamma=0.55+0.05[1+w(z=1)]$, 
\cite{groexp} or simply $f=\om(a)^{0.55}$ for a $\Lambda$CDM universe. 
(And as shown in Fig.~\ref{fig:betasens}, ignoring the $w$ dependence of 
$\gamma$ is a good approximation.) 

\begin{figure}[!htb]
\begin{center}
\psfig{file=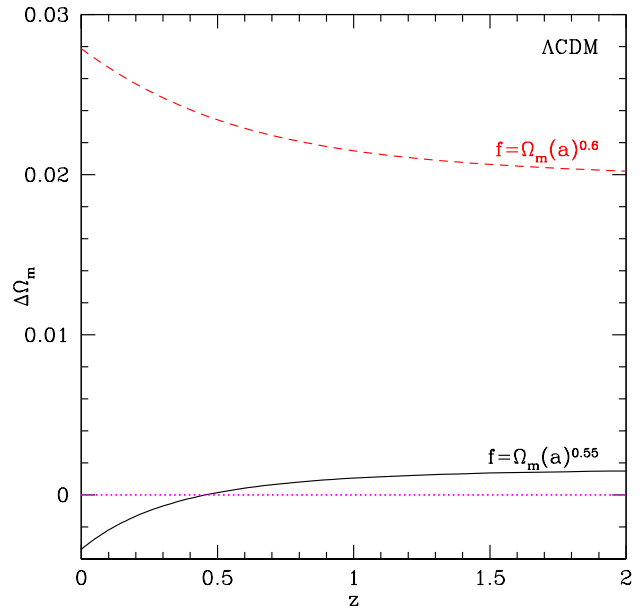,width=3.4in}
\caption{Using redshift distortions to measure the matter density requires 
the proper form for the growth factor.  The 1980 form (before CDM and 
dark energy) with growth index of $0.6$ will misestimate $\om$ by up 
to 0.03, while the modern form with $0.55$ is more than 10 times as accurate. 
}
\label{fig:ombias}
\end{center}
\end{figure}

Figure~\ref{fig:ombias} illustrates the bias induced in the derived 
matter density if the incorrect form for the growth factor is employed. 
We see that using $f=\om(a)^{0.55}$ is at least 10 times more accurate 
than $f=\om(a)^{0.6}$.  (The form of \cite{lahav} improves on the $0.6$ 
form by not quite a factor two; similar results on relative accuracy of the 
forms hold for dark energy other than $\Lambda$.) 
Using $f=\om(a)^{0.6}$ can bias the value 
of $\om$ derived from the measurements by $\sim0.03$, a substantial offset 
for precision cosmology.  By contrast, $f=\om(a)^{0.55}$ recovers $\om$ 
to better than 0.003, with zero error relative to the exact numerical 
solution for growth in the important $z\approx0.5$ range.

\section{Conclusion \label{sec:concl}}

Testing gravity is an important experimental goal at the intersection 
of astrophysics and fundamental physics.  The development of redshift 
distortion measurements in galaxy redshift surveys is an exciting advance. 
We have shown that 10\% measurements from $z=0-1$, well within future 
observational capabilities, in complementarity with independent, accurate 
distances from $z=0-1.7$ to measure the expansion history, can put 
significant limits on gravitational modifications.  
The growth and expansion methods work in parallel to test both gravity and 
dark energy. 

Redshift distortions 
have little leverage on the dark energy equation of state but can determine 
the matter density accurately if the modern form $f=\om(a)^{0.55}$ is used. 

While issues of galaxy bias and the exact relation between 
the velocity and matter power spectra (e.g.\ \cite{kaiser,scocci}) add some 
uncertainty, further observational leverage exists also.  According to 
\cite{guzzo}, improved knowledge of the galaxy spatial correlation 
function could reduce error bars a factor of three, while larger surveys, 
such as extending the Vimos VLT Deep Survey (VVDS) to 400 deg$^2$, could 
estimate $\beta$ to better than 5\%. 

Redshift distortions are exciting 
especially because of the strong motivation for testing gravity on these 
length scales.  While cosmological observations probe the theory of 
gravity on 1000 Mpc scales, and galaxy dynamics on 1-100 kpc scales, 
velocity flows give insight on 1-30 Mpc scales where many theories predict 
a transition from the form causing the cosmic acceleration on large scales 
to the general relativity limit on small scales.  As pointed out by 
\cite{pjep02}, the relation between the velocity field of matter 
and the growth of density perturbations also probes the inverse square law 
of gravity, while \cite{zhang} has emphasized that the redshift distortions 
have a different dependence on the metric potentials than do gravitational 
lensing shape distortions.  Measuring redshift distortions, and in 
particular a value of $\gamma$ different from the Einstein value of 0.55, 
could thus point the way to a new understanding of gravitation.

\acknowledgments 

I thank Gigi Guzzo for motivating this investigation and the Aspen Center 
for Physics for providing the venue for such discussions. 
This work has been supported in part by the Director, Office of Science, 
Department of Energy under grant DE-AC02-05CH11231.

\end{document}